\documentclass[epj]{svjour}
\usepackage{graphics}

\def\bmatrix{\left[\begin{array}}
\def\ematrix{\end{array}\right]}

\newcommand{\gtap}{ \raisebox{-0.6ex}
                       {$\ \textstyle \stackrel{\textstyle >}{\sim}\ $}
                      }
\begin{document}
\title{Effects of New Physics on CP Violation in B Decays}
\author{George W.S. Hou 
}                     
%
%
\institute{Department of Physics, National Taiwan University,
Taipei, Taiwan 10764 
}
\date{Received: date / Revised version: date}
%
\abstract{
We discuss two models with 1 extra CP phase in $b\leftrightarrow
s$ transition.
The CP phase $\arg (V_{t^\prime s}V_{t^\prime b})$ with fourth
generations, previously ignored, could impact on $b\to
s\ell^+\ell^-$, $\Delta m_{B_s}$ and $\sin2\Phi_{B_s}$, but does
not affect EM and strong penguins.
With SUSY at TeV scale, a right-handed ``$\widetilde{sb}_{1}$''
squark can be driven light by flavor mixing. It does not affect
$b\to s\ell^+\ell^-$, but can generate $S_{\phi K_S} < 0$ while
giving $S_{\eta^\prime K_S} \sim \sin2\Phi_{B_d} \cong0.74$. $B_s$
mixing and $\sin2\Phi_{B_s}$ would likely be large, and
$S_{K_S\pi^0\gamma}\neq 0$ in $B^0\to K^{*0}\gamma$ is promising.
} 
\maketitle
\section{Introduction}
\label{intro}

With $\sin2\Phi_{B_d}$ agreeing with CKM fit, New Physics (NP)
seems absent in $B_d$ mixing, but $b\leftrightarrow s$ transitions
seem fertile. The large $K\pi/\pi\pi$ ratio shows the importance
of penguins.
More intriguing~\cite{Browder} is the hint of $S_{\phi K_S} < 0$,
although $S_{\eta^\prime K_S} \sim \sin2\Phi_{B_d}$. Belle's 2003
result~\cite{Browder,phiKs} is 3.5$\sigma$ from 0.74. Despite
BaBar's change in sign, this is still a strong indication for NP
in $b\to s$ penguins.

$B_s$ mixing has been ``just around the corner" since the 1990s.
It eliminates the second quadrant for $\phi_3$ in the CKM fit, but
this would no longer hold if NP lurks. The litmus test for NP
would be to find $\sin2\Phi_{B_s} \neq 0$, hopefully in the near
future. Another clear sign for NP would be wrong helicity photons
in $b\to s\gamma_L$, which can be tested via measuring $S_{B_s\to
\phi\gamma}$, or by measuring $\Lambda$ polarization in $\Lambda_b
\to \Lambda\gamma$. However, there is now hope to
reconstruct~\cite{Browder} $B_d$ vertex from $K_S$ at B factories,
allowing one to measure $S_{B_d\to K_S\pi^0\gamma}$ where
$K_S\pi^0$ comes from $K^{*0}$.

The present is already bright for NP search in $b\leftrightarrow
s$ transitions, and the future can only be brighter! To elucidate
the possibilities lying ahead for us, we focus on models that
bring in {\it just 1 extra CP phase} in $b\leftrightarrow s$.
The first model is that of a 4th generation~\cite{AH}.
The second is large $\tilde s_R$-$\tilde b_R$
mixing~\cite{ACH,CHN} with SUSY.
%

\section{4th Generation} \label{4th}

It is peculiar that, since the early~\cite{HWS} discussions of
impact of 4th generation on $b\to s\ell\ell$, where
$\lambda_{t^\prime} \equiv V_{t's}^*V_{t'b} \equiv r_s \, e^{i
\Phi_s}$ was taken as real for convenience, the literature that
followed mostly ignored the possibility of $\Phi_s\neq 0$.

It is true that $\lambda_t \cong -\lambda_c-\lambda_{t^\prime}
\cong -0.04-\lambda_{t^\prime}$for $r_s = \vert
\lambda_{t^\prime}\vert \gg \vert \lambda_{u}\vert \approx
\lambda^5 \sim 0.0006$. For a typical operator $O_i(\mu)$, its
coefficient is changed from $\lambda_{t} C_i^{\rm SM}(\mu)\to
\lambda_{t} C_i^{\rm SM}(\mu) + \lambda_{t'} C_i^{\rm new}(\mu)$.
By simple rearrangement one gets,
\begin{eqnarray}
\lambda_t C_{i}^{\rm SM}+\lambda_{t^\prime}C_{i}^{\rm new} =
-\lambda_c C_{i}^{\rm SM} +\lambda_{t'} (C_{i}^{\rm new}
-C_{i}^{SM}), \label{decomp}
\end{eqnarray}
where {\it the first term is the usual SM contribution}. The
second term is the genuine 4th generation effect. It vanishes for
$m_{t'}\to m_t$ or $\lambda_{t'} \to 0$, as required by GIM. What
has been popular, besides ignoring $\Phi_s$, is to absorb
$\lambda_{t^\prime}$ into the definition of $C_{i}^{\rm new}$.
This is rather bad practice.

We have 3 new parameters, $m_{t^\prime}$, $r_s$ and $\Phi_s$,
where we are most interested in the latter. The constraints from
${\cal B}^{\rm expt}(B\to X_s \gamma) = (3.3\pm 0.4)\times
10^{-4}$, which agrees with SM3, is rather weak. $B_s$ mixing is
strongly dependent on $m_{t^\prime}$. Choosing SM parameters such
that $\Delta m_{B_s}^{\rm SM3} = 17.0\ {\rm ps}^{-1}$, the bound
of $14.9\ {\rm ps}^{-1}$ disfavors $0\leq r_s\leq 0.03$ and
$\cos\Phi_s>0$, because $t^\prime$ effect is destructive. The
allowed parameter space is larger for lower $m_{t'}$, but the most
forgiving zone is when $\Phi_s \sim \pi/2$ or $3\pi/2$, i.e.
purely imaginary, when $t^\prime$ effects add in quadrature to
SM3!

One interesting test ground for SM4 is $b\to
s\ell\ell$~\cite{HWS}, since the EW or Z penguin has strong
$m_{t^\prime}$ dependence like $\Delta m_{B_s}$. Unlike $\Delta
m_{B_s}$, however, several modes are now measured. The first
measurement of $B\to K\ell\ell$ was on the high side of SM3, which
motivated our study of SM4~\cite{AH}. Now the number has come
down, and both $B\to K\ell\ell$ and $K^*\ell\ell$ are not in
disagreement with SM.

\begin{figure}
\resizebox{0.46\textwidth}{!}{%
\hskip1cm
  \includegraphics{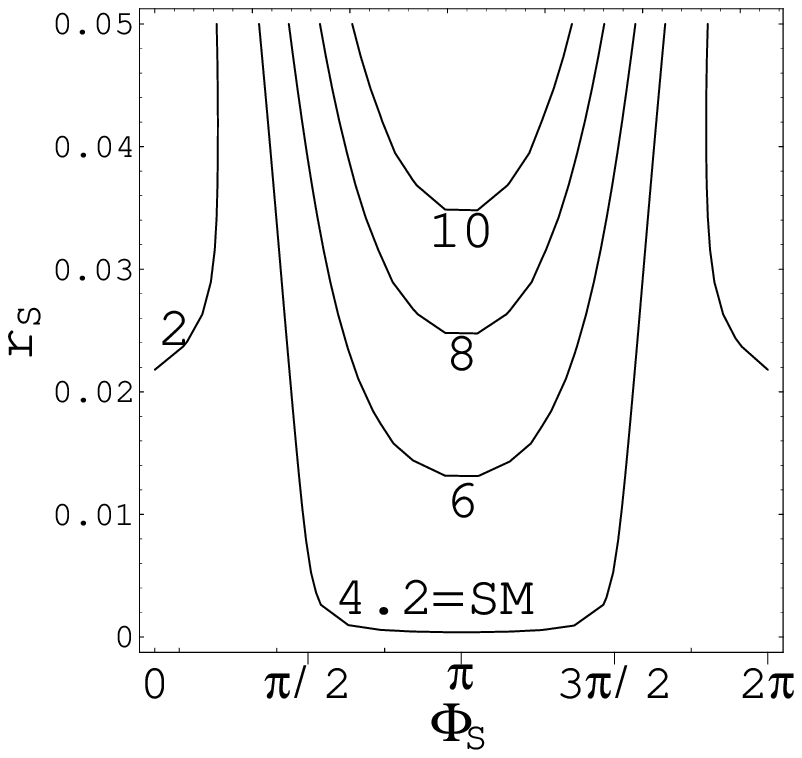} \hskip0.3cm
  \includegraphics{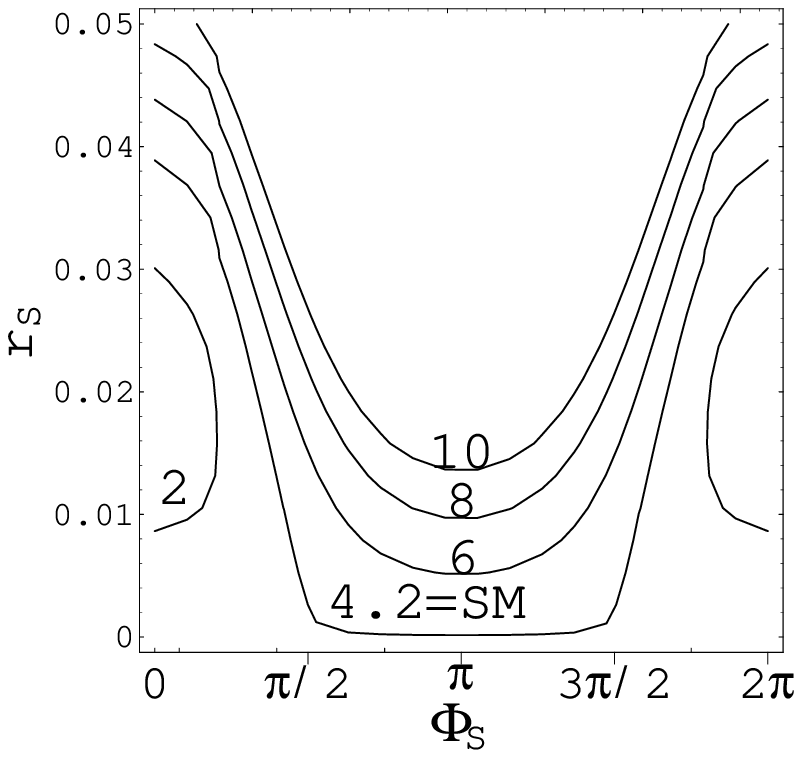}
}
\caption{${\cal B}(B\to X_s\ell^+\ell^-)\times10^6$ for
$m_{t^{\prime}}=$ (a) 250, (b) 350 GeV.}
\label{fig:sll}       
\end{figure}

In any case, the exclusive rates have larger hadronic
uncertainties, so let us focus on the inclusive, where the current
Belle result of ${\cal B}(B\to X_s\ell^+\ell^-) =
(6.1\pm1.4^{+1.3}_{-1.1}) \times 10^{-6}$ is slightly higher than
SM3 expectation of $\sim 4.2\times 10^{-6}$, partly because NNLO
result dropped by 40\%. In Fig.~\ref{fig:sll} we plot ${\cal
B}(B\to X_s\ell^+\ell^-)$ contours in $\Phi_s$-$r_s$ plane, for
$m_{t^\prime} =$ 250 and 350 GeV. For $\cos\Phi_s>0$, $B\to X_s
l^+l^-$ is less than $4.2\times 10^{-6}$ hence less favored. The
behavior for $\pi/2 < \Phi_s < 3\pi/2$ is rather similar to
$\Delta m_{B_s}$, but provides more stringent bounds since $B_s$
mixing is not yet measured. Furthermore, it will more readily
improve. In a way, one may say that if NNLO result for SM3 remains
low, if refined experiment still gives $5\times 10^{-6}$, SM4 may
be called for. Again we note that $\Phi_s \sim \pi/2$ or $3\pi/2$
is more accommodating, and allows for larger $r_s$. However, there
is no further information in $m^2_{\ell\ell}$ spectrum, and,
constrained by the observed rate, ${\cal A}_{FB}$ is as in SM3.

%

\begin{figure}[b]
\resizebox{0.46\textwidth}{!}{%
\hskip0.7cm
  \includegraphics{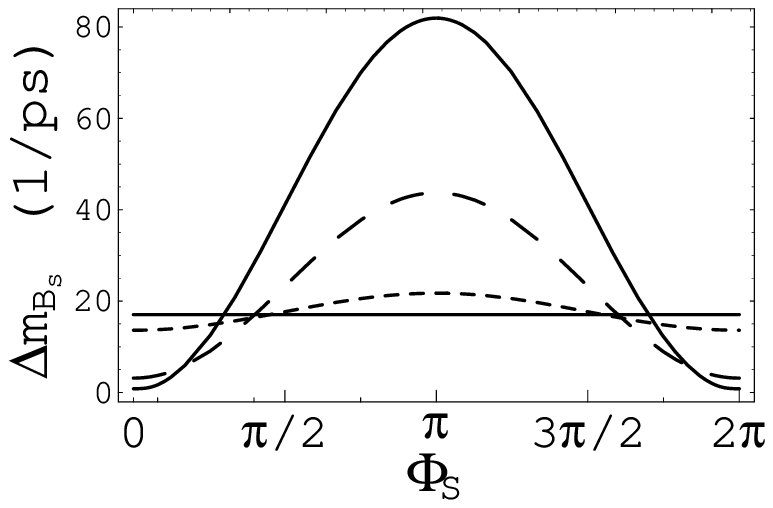} \hskip0.2cm
  \includegraphics{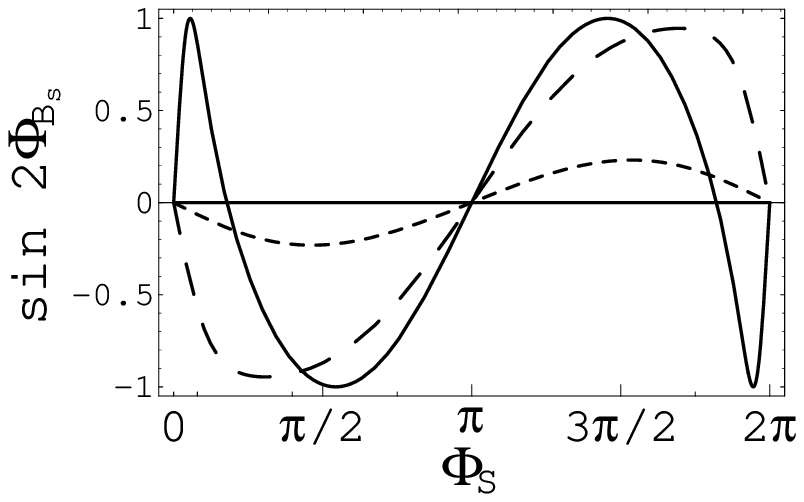}
}
\resizebox{0.46\textwidth}{!}{%
\hskip0.7cm
  \includegraphics{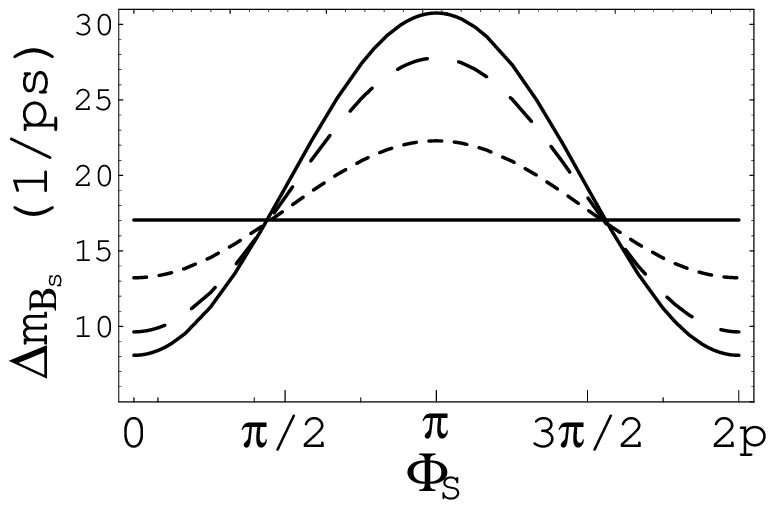} \hskip0.2cm
  \includegraphics{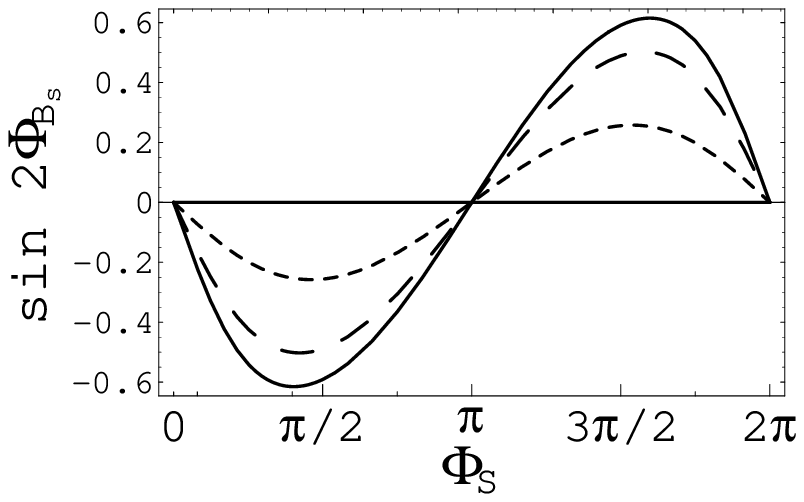}
}
\caption{$\Delta m_{B_s}$ and $\sin 2\Phi_{B_s}$ vs. $\Phi_s$.
Short dash, long dash and solid lines for upper (lower) plots are
for $r_s=$ 0.002, 0.01, 0.02 (0.002, 0.004, 0.005) and for
$m_{t^{\prime}}=$ 250~(350)~GeV.}
\label{fig:Bs}       
\end{figure}

The highlight for SM4, by considering CP phase $\Phi_s$, is
prospect for sizable $\sin2\Phi_{B_s}$, where any nonvanishing
value would indicate NP.
We define $\Delta m_{B_s} = 2 |M_{12}|$ and
$M_{12}=|M^B_{12}|e^{i\Phi_{B_S}}$. As the box diagrams can
contain none (SM3), one or two $t^\prime$ legs, we have
\begin{eqnarray}
M_{12}=|M_{12}| e^{2 i \Phi_{B_s}} \approx  r_s^2 e^{2 i \Phi_s} A
+ r_s e^{i\Phi_s} B + C
 \label{M12Phis}
\end{eqnarray}
where $A$ and $B$ are explicit functions of $m_t$ and $m_{t'}$ and
$C$ is the usual SM3 contribution. This allows us to understand
the change of ``periodicity" of $\sin 2\Phi_{B_s}$ vs. $\Phi_s$ in
Fig.~\ref{fig:Bs}, which plots both $\Delta m_{B_s}$ (left) and
$\sin 2\Phi_{B_s}$ for $m_{t^{\prime}}=$ 250, 300~GeV for several
$r_s$ values. The straight lines are the SM3 expectations. For
$\Delta m_{B_s}$ this is slightly above experimental bound. Thus,
only the $\Phi_s$ range where $\Delta m_{B_s}$ falls a little
below the straight line is ruled out.

We offer several observations on prospects for $\sin 2\Phi_{B_s}$
by inspection of Fig.~\ref{fig:Bs}: (1) Even small $r_s$ values
can give sizable $\sin 2\Phi_{B_s}$; (2) Both signs are possible;
(3) Largest if $\Delta m_{B_s}$ is ``just around the corner", i.e.
to be measured soon.
This last point makes SM4 very interesting at the Tevatron Run II.
As discussed, $\Delta m_{B_s}$ hovers around SM3 expectation for
$\Phi_s \sim \pi/2$ or $3\pi/2$, when all constraints are most
accommodating because they add in quadrature to SM3 effects, {\it
except} in the direct measure of CP phase, $\sin 2\Phi_{B_s}$. One
has the ideal situation that $\Delta m_{B_s}$ is most measurable,
while $\sin 2\Phi_{B_s}$ can vary between $\pm 1$.

\section{Light \boldmath $\widetilde{sb}_{1R}$ Squark}
\label{sb}

The 4th generation is not effective on EM and strong penguins,
because $t$ and $t^\prime$ effects are very soft for such loops.
Furthermore, the chirality is the same as SM3, i.e. left-handed,
hence only the usual right-handed helicity photons appear in $b\to
s\gamma$. The scenario of a light $\widetilde{sb}_{1R}$ squark,
however, can touch all these aspects as well as $B_s$ mixing,
though it does not affect $b\to s\ell\ell$.

Large $\tilde s_R$-$\tilde b_R$ mixing can be related, in the
context of SUSY-GUT, to~\cite{HLMP} the observed near maximal
$\nu_\mu$-$\nu_\tau$ mixing. While this is attractive in itself,
we prefer not to assume the behavior at high scale, but to look at
what data demands. The 2003 average for $S_{\phi K_S} = -0.15\pm
0.33$ is still 2.7$\sigma$ from SM expectation of 0.74. As this
would be a large NP $b\to s$ CP violation effect, it would demand
{\it (i) large effective $s$-$b$ mixing}, and the presence of a
{\it (ii) large new CP phase}. Furthermore, to allow for
$S_{\eta^\prime K_S} \sim \sin 2\Phi_{B_d}$, the {\it (iii) new
interaction should be right-handed}~\cite{KhaKou}. We find it
extremely interesting that all three aspects are brought about
{\it naturally} by the synergies of {\it Abelian} flavor symmetry
(AFS) and SUSY. We will see that AFS brings in large $s_R$-$b_R$
mixing, and SUSY makes this dynamical, and also activating one new
CP phase in $\tilde s_R$-$\tilde b_R$ mixing.

Focusing only on the 2-3 down sector, the normalized $d$ quark
mass matrix has the elements $\hat M^{(d)}_{33} \simeq 1$, $\hat
M^{(d)}_{22} \simeq \lambda^2$, while taking analogy with $V_{cb}
\simeq \lambda^2$ gives $\hat M^{(d)}_{23} \simeq \lambda^2$. But
$\hat M^{(d)}_{32}$ is unknown for lack of right-handed flavor
dynamics. With effective AFS~\cite{Nir}, however, the {\it
Abelian} nature implies $\hat M^{(d)}_{23}\hat M^{(d)}_{32} \sim
\hat M^{(d)}_{33}\hat M^{(d)}_{22}$, hence $\hat M^{(d)}_{32} \sim
1$ is deduced. This may be the largest off-diagonal term, but its
effect is hidden within SM.
With SUSY, the flavor mixing extends to $\tilde s_R$-$\tilde b_R$,
which we parametrize as
%
\begin{equation}
\widetilde M^{2(sb)}_{RR}  = \left[
\begin{array}{ll}
\widetilde m_{22}^2 &   \widetilde m_{23}^2 e^{-i\sigma} \\
\widetilde m_{23}^2 e^{i\sigma}  &  \widetilde m_{33}^2
\end{array}  \right]
  \equiv R \left[
    \begin{array}{cc}
    \widetilde m_{1}^2 & 0 \\
    0 & \widetilde m_{2}^2
    \end{array}  \right]
    R^\dagger,
%
 \label{sbsquark}
\end{equation}
where $\widetilde m^2_{ij} \simeq \widetilde m^2$, the common
squark mass, and
\begin{eqnarray}
R = \bmatrix{cc} \cos\theta & \sin\theta \\
-\sin\theta e^{i\sigma} & \cos\theta e^{i\sigma} \ematrix.
 \label{R}
\end{eqnarray}
There is {\it just one}~\cite{ACH} CP phase $\sigma$, which is on
equal footing with the KM phase $\delta$ as both are rooted in the
quark mass matrix.
Note that $\widetilde M^2_{LR} = (\widetilde M^2_{RL})^\dagger
\sim \widetilde{m} M$ is suppressed by quark mass, while
$\widetilde M^2_{LL}$ is CKM suppressed.

\begin{figure}[t]
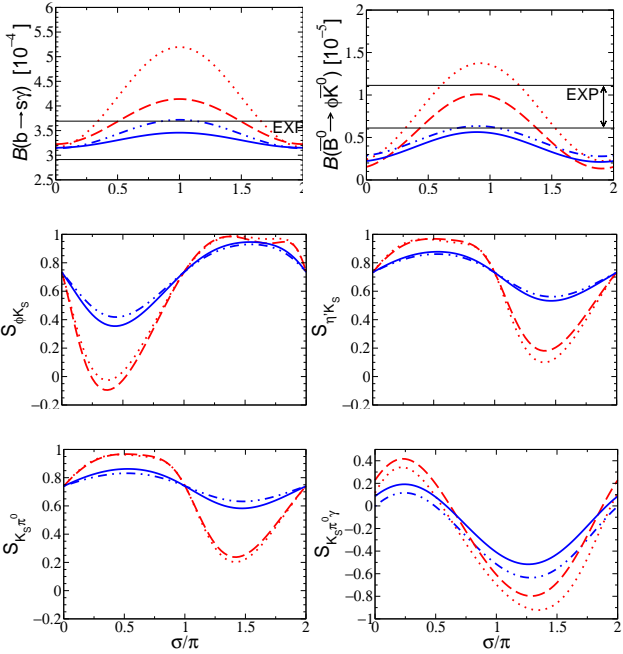

\resizebox{0.46\textwidth}{!}{%
\hskip0.7cm
  \includegraphics{bsp_br.eps} \hskip0.4cm
  \includegraphics{phik_br.eps}
} \vskip0.2cm
\resizebox{0.46\textwidth}{!}{%
\hskip0.3cm
  \includegraphics{sphik.eps} \hskip0.2cm
  \includegraphics{setak.eps}
} \vskip0.4cm
\resizebox{0.463\textwidth}{!}{%
\hskip0.5cm
  \includegraphics{sk0pi0.eps} \hskip0.2cm
  \includegraphics{mix_f.eps}
}
\caption{(a) ${\cal B}(b\to s\gamma)$, (b) ${\cal B}(B^0\to\phi
K^0)$, (c) $S_{\phi K_S}$, (d) $S_{\eta^\prime K_S}$, (e)
$S_{K_S\pi^0}$ and (f) $S_{K_S\pi^0\gamma}(\bar B^0\to \bar
K^{*0}\gamma)$ vs $\sigma$ for $\widetilde m_1 = 200$ GeV and
compared with experiment. Solid, dotdash (dash, dots) lines are
for $\widetilde{m} =$ 2, 1 TeV, $m_{\tilde g}=$ 0.8 (0.5) TeV.}
\label{fig:sbresults}       
\end{figure}

The presence of large flavor violation in squark masses pushes
SUSY scale to above TeV, even after one decouples
$d$-flavor~\cite{ACH}. Interestingly, the near democratic nature
of Eq. (3) allows, by some fine tuning, one squark to be driven
light by the large mixing. We denote this squark
$\widetilde{sb}_{1R}$, and take its mass at 200 GeV for
illustration (so $\widetilde{sb}_{2R}$ would have mass
$2\widetilde m^2$). The presence of right-handed $s_R\tilde
b_R\tilde g$ couplings doubles the operators $O_i$ by flipping
chirality, to $O_i^\prime$. We calculate coefficients $c_i$ and
$c_i^\prime$ in mass basis, and evaluate matrix elements in naive
factorization. The most interesting effect occurs to photonic and
gluonic dipole penguins, in particular $c_{11}^\prime$ and
$c_{12}^\prime$. Let us now just discuss the salient results,
which are plotted in Fig.~\ref{fig:sbresults}.

Fig.~\ref{fig:sbresults}(a) shows that $b\to s\gamma$ is rather
accommodating. This is because the right-handed effect adds only
in quadrature to $b\to s\gamma$ rate~\cite{ACH}.
We cannot account for $B\to\eta^\prime K$ rate, but
Fig.~\ref{fig:sbresults}(b) shows that $B\to \phi K_S$ rate can in
principle be brought up for $\cos\sigma < 0$, while it is known
that the standard gluonic dipole penguin ($c_{12}$) suppresses the
rate. It is amusing that if one takes the two rates together as
constraints, purely imaginary $\sigma$ is preferred, which is
further born out from CP measurables.

Fig.~\ref{fig:sbresults}(c) plots the enigmatic $S_{\phi K_S}$ vs.
$\sigma$. It is interesting that the low $\widetilde{sb}_{1R}$
mass, together with a low $m_{\tilde g}$ mass of 500 GeV,
can~\cite{HLMP,KhaKou} bring $S_{\phi K_S}$ negative for $\sigma
\sim \pi/2$. However, as seen from Fig.~\ref{fig:sbresults}(d),
$S_{\eta^\prime K_S}$ stays above $\sin2\Phi_{B_d} \cong 0.74$
hence is positive~\cite{KhaKou}. This is due to right-handed
interactions. More specifically, one has
\begin{eqnarray}
{\cal A}(\bar B^0\to \phi \bar K^0)
 \propto \Bigl\{\cdots
 +\frac{\alpha_s}{4\pi}\frac{m_b^2}{q^2} \, \tilde{S}_{\phi K} \,
(c_{12}+c_{12}^\prime)\Bigr\},
 \label{AphiK}
\end{eqnarray}
where $\cdots$ are several terms $\propto a_i + a_i^\prime$, and
${\cal A}(\bar B^0\to \eta^\prime \bar K^0)$ is even more
complicated, but the crucial point is a sign change for the
$c_{12}^\prime$ term. Pseudoscalar production picks up the sign of
the axial current!

Besides elucidating how $S_{\phi K_S} < 0$ while $S_{\eta^\prime
K_S} \sim \sin2\Phi_{B_d}$ can be maintained, Eq. (5) also shows
the elements in enhancing the effect of $c_{12}^\prime$. Lowering
squark and gluino masses enhances $c^\prime_{12}$, but we also
have the hadronic parameters $\tilde S_{\phi K}/q^2$. We resort to
these for further enhancement rather than lowering $m_{\tilde g}$
further.

Having zoomed into $\sigma \sim 65^\circ$ as ``preferred", we were
surprised to find, contrary to our earlier thought~\cite{ACH},
that the lighter gluino makes $\Delta m_{B_s} \gtap 70$ ps$^{-1}$
rather difficult to avoid~\cite{CHN}, even though $\sin
2\Phi_{B_s}$ could vary through 0 to 1. Reminded by the sluggish
start of Tevatron Run~II, it seems that $\Delta m_{B_s} \gtap 70$
ps$^{-1}$ would have to await LHCb or BTeV. What is worse, even
with $\Delta m_{B_s}$ measured some years from now, the very fast
$B_s$ oscillations would make, with the exception of perhaps $\sin
2\Phi_{B_s}$ itself, much of the CP program in $B_s$ decay rather
difficult.

We are, however, intrigued by a very recent development. BaBar has
made a first attempt~\cite{Browder} at measuring $S_{K_S\pi^0}$,
``reconstructing" the $B^0$ vertex by extrapolating $K_S$ momentum
onto the boost, i.e. $B$ direction, a knowledge that is unique to
B factories. They find $S_{K_S\pi^0} = 0.48^{+0.38}_{-0.47}\pm
0.10$, which is in agreement with our results shown in
Fig.~\ref{fig:sbresults}(e). The features are similar to
$S_{\eta^\prime K_S}$ of Fig.~\ref{fig:sbresults}(d), since both
are $PP$ final states.
What excites us is the prospect for measuring mixing dependent CP
in $B\to K^{*0}\gamma$, formerly thought impossible, but now
hopeful with this ``$K_S$ vertexing" technique. We note that
\begin{equation}
{S}_{M^{0}\gamma} = \frac{2\vert c_{11}c^\prime_{11}\vert}{\vert
c_{11}\vert^2 + \vert c^\prime_{11}\vert^2} \,\xi\,
\sin\left(2\phi_{B_d}-\varphi_{11}-\varphi_{11}^\prime\right),
\end{equation}
where $\xi$ is the CP of reconstructed $M^0$ final state, and
$\phi_{B_d} = \phi_1$, $\varphi_{11}^{(\prime)}$ are the phases of
$B_d$ mixing and $c_{11}^{(\prime)}$, respectively. For $B\to
K^{*0}\gamma$ with $K^{*0}$ decaying to CP eigenstate $K_S\pi^0$,
Eq.~(6) can be completely calculated, with little hadronic
uncertainty, which we plot in Fig.~\ref{fig:sbresults}(f).

{\it The finiteness of this single measurable justifies the
luminosity upgrades of B factories}, currently being contemplated,
because it provides a clean measure and confirmation of the type
of NP. The measurables such as $S_{\phi K_S}$, $S_{\eta^\prime
K_S}$ and $S_{K_S\pi^0}$, tantalizing as they might be, are
plagued by hadronic parameters such as $\tilde S_{\phi K}/q^2$. We
note, finally, that $S_{K_S\pi^0\gamma}(\bar B^0\to \bar
K^{*0}\gamma)$ is close to impossible to measure at hadronic
machines, for not knowing $B$ direction, and having too many
photons.

SuperB upgrades should invest on a large Silicon Vertex Detector.


%
%

\end{document}